\documentclass[pra,aps,twocolumn,showpacs]{revtex4}
\usepackage{graphicx}
\usepackage{dcolumn}

\begin{document}
\def\be{\begin{equation}}
\def\ee{\end{equation}}
\def\bearr{\begin{eqnarray}}
\def\eearr{\end{eqnarray}}
\def\la{\langle}
\def\ra{\rangle}
\def\l{\left}
\def\r{\right}

\title{Exact solutions for a Dirac electron in an exponentially 
decaying magnetic field}
\author{Tarun Kanti Ghosh}
\affiliation{Department of Physics, Indian Institute of Technology-Kanpur, 
Kanpur 208 016, India }
\date{\today}

\begin{abstract}
We consider a Dirac electron in the presence of an exponentially 
decaying magnetic field. We obtain exact energy eigenvalues with 
a zero-energy state and the corresponding eigenfunctions.
We also calculate the probability density and current distributions.
\end{abstract}
%Dirac eq. 03.65.Pm,Graphite, 81.05.Uw,Landau levels, 71.70.Di,
%two-dimensional electron gas 73.20.-r, 
%Surface physics, nanoscale physics, low-dimensional systems

\maketitle

The experimental realization of graphene, two-dimensional (2D) sheet
of graphite \cite{dis,dis1}, and of the massless Dirac nature of its 
electron low-energy spectrum \cite{dirac,dirac1} has given rise to a 
tremendous interest in this field (for recent reviews, see 
\cite{review,review1}).
The energy spectrum which goes linearly with the momentum and the specific 
density of states of the Dirac electrons \cite{wallace} enabled the study 
of experimentally chiral tunneling and the Klein paradox in graphene 
\cite{klein}. 
This also leads to the anomalous Landau level spectrum in a uniform magnetic 
field, which gives rise to the half-integer quantum Hall effect \cite{dirac,qhe}.   

The discovery of the half-integer quantum Hall effect and the zero-energy
Landau level \cite{qhe} stimulated lot of theoretical research interest on 
Dirac electrons in uniform as well as non-uniform magnetic fields. 
For example, the Dirac-Weyl equation has been solved numerically for a 
single electron in a step-like magnetic field, magnetic barrier and 
polytropic magnetic field with $ B = B_0 y^{\gamma} $, where $ \gamma > 0 $ 
\cite{egger1,lambert,egger2,egger3}.
We should also mention that the Schr${\ddot {\rm o}}$dinger equation has been solved 
numerically 
for spinless 2D electrons in a linearly varying magnetic field \cite{mueller}, 
step-like magnetic field, magnetic barrier \cite{peeters,peeters1,sch}. 
An analytical solution has been given for an electron in the presence of 
an exponentially decaying magnetic field \cite{exp}.

In this work, we solve analytically for the Dirac electron in graphene in 
the presence of an exponentially decaying magnetic field 
$B = B_0 e^{-\lambda x}$. 

The Hamiltonian of the massless electrons in graphene near one of the
Dirac $ K $ points is described by a two-component Dirac-Weyl 
equation
\be
H = v_F {\bf \sigma} \cdot {\bf p},
\ee
where $ v_F \approx 10^6$~m/sec is the Fermi velocity,
$ {\bf \sigma } = \{\sigma_x, \sigma_y\} $ is the $ 2 \times 2 $ 
Pauli matrices and $ {\bf p} = -i \hbar {\boldmath \nabla} $ is the 
two-dimensional momentum operator. The pseudo-relativistic 
dispersion relation with Fermi velocity arises is due to the 
sublattice structure: the basis of the graphene honeycomb lattice 
contains two carbon atoms, giving rise to an isospin degree of freedom.
In the presence of an external magnetic field 
($ {\bf B} = {\bf \nabla} \times {\bf A} $) 
perpendicular to the graphene plane, the Hamiltonian of the single Dirac 
electron $(-e)$ is 
$ H =  v_F {\bf \sigma} \cdot ({\bf p} + e {\bf A}) $.

We consider an electron in the graphene sheet in the presence of a 
non-uniform magnetic field 
$ {\bf B}(x,y) = \{0, 0, B_0 e^{-\lambda x} \hat z \}$ perpendicular to 
the $x-y$ plane. In the Landau gauge, 
the corresponding vector potential is 
$ {\bf A}(x,y) = \{0, -(1/\lambda) B_0 e^{-\lambda x} \hat y, 0\}$.
When $ \lambda \rightarrow 0 $, the non-uniform magnetic field becomes 
a constant magnetic field.
The time-independent Dirac-Weyl equation is
\begin{equation} \label{meq}
v_F {\bf \sigma} \cdot ({\bf p} + e {\bf A}) \Psi(x,y) = E \Psi(x,y).
\end{equation}
Here, $ \Psi (x,y) = \{\Psi_+(x,y), \Psi_-(x,y)\}^T $ is the two-component 
wavefunction and $ T $ denotes the transpose of the column vector. 
Due to the translation  invariance in the $y$-direction, the longitudinal
momentum $ k_y $ is a conserved quantity. One can parameterize the 
wavefunction as $\Psi(x,y)= e^{ ik_y y} \{\psi_+(x), \psi_-(x)\}^T $.
From equation (\ref{meq}), one can get two coupled equations for 
$ \psi_+ $ and $ \psi_-$ as given below:

\begin{equation} \label{dir1}
-i \hbar v_F [ \frac{\partial}{\partial x} + (k_y + \frac{e}{\hbar} A_y)] 
\psi_- (x) = E \psi_+(x)
\end{equation}
and
\begin{equation}\label{dir2}
-i \hbar v_F [ \frac{\partial}{\partial x} - (k_y + \frac{e}{\hbar} A_y)]
\psi_+ (x) = E \psi_-(x).
\end{equation}

Using two coupled equations. (\ref{dir1}) and (\ref{dir2}), we get 
Schr${\ddot {\rm o}}$dinger-like 
decoupled equations for $ \psi_+ $ and $ \psi_-$  as 

\begin{equation} \label{dirac1}
[- (\hbar v_F)^2 \frac{\partial^2}{\partial x^2} + V_+(x) ] \psi_+ = E^2 \psi_+
\end{equation}
and
\begin{equation} \label{dirac2}
[- (\hbar v_F)^2 \frac{\partial^2}{\partial x^2} + V_-(x) ] \psi_- = E^2 \psi_-.
\end{equation}
Here, the effective potentials $ V_{\pm} $ are given by 
\begin{equation} \label{potential}
V_{\pm} = (\hbar v_F)^2 [\pm \frac{e}{\hbar} \frac{\partial }{\partial x} A_y(x) + 
(k_y + \frac{e}{\hbar} A_y)^2]. 
\end{equation}
Here, the first term on the right hand side of the above equation is 
the Zeeman-like term due to the isospin degree of freedom in presence 
of the magnetic field. Define the magnetic length scale as 
$ l(x) = \sqrt{\hbar/e B(x)}$.

As seen in figure 1, the effective potentials $ V_{\pm} (x)$ have the
form of an asymmetric quantum wells formed by the exponentially decaying
magnetic field. It is well known that such a well can have a bound state 
if the well is sufficiently deep. Figure 1 will be discussed in more 
detail later on.

Following the references. \cite{exp,landau}, we introduce two new 
dimensionless variables:
\begin{equation}
\xi(x) = \frac{1}{ (l(x) \lambda)^2} = \frac{eB_0 e^{-\lambda x}}{\hbar \lambda^2}
\end{equation}
and
\begin{equation}
\xi_0 = \frac{|k_y|}{\lambda} \equiv  \frac{e B_0 e^{-\lambda x_0}}{\hbar \lambda^2}
= \frac{1}{(l(x_0) \lambda)^2}. 
\end{equation}
Here, $ l(x_0) = \sqrt{\hbar/e B(x_0)} $ is the magnetic
length scale for a given value of $ x = x_0 $ which depends on the 
conserved wavevector $ k_y$ through equation (8). Also, $ \xi $ varies 
from $ 0 $ to $ \infty $ when $ x $ varies from $ \infty $ to 
$ - \infty $. 

In the new variables, equations. (\ref{dirac1}) and (\ref{dirac2}) 
reduce to
\begin{equation}
[\frac{d^2}{d \xi^2} + \frac{1}{\xi} \frac{d}{d \xi} -\frac{\beta^2}{\xi^2} +
\frac{2\xi_0 \mp 1}{\xi} -1 ] \psi_{\pm} = 0, 
\end{equation} 
where $ \beta ^2 = \xi_0^2 - \epsilon^2 $ and $ \epsilon = E/(\hbar v_F \lambda)$.
The behavior for small and large $ \xi $ suggests that the general 
solution can be written as 
$ \psi_{\pm} (\xi) \sim \xi^{\beta} e^{-\xi} w_{\pm}(\xi) $. Inserting this ansatz
in the previous equation, we get for $ w_{\pm} (z = 2 \xi) $ 
\begin{equation}
[z \frac{d^2}{dz^2} + (\gamma -z)  \frac{d}{dz} - \alpha_{\pm} ]w_{\pm}(z) = 0 
\end{equation} 
which has the form of a confluent hypergeometric equation, where 
$ \gamma = 2 \beta + 1 $ and 
$ \alpha_{\pm} = - \xi_0 + \beta  + \frac{1}{2} \pm \frac{1}{2} $.
Two linearly independent solutions can be chosen as 
$ F[\alpha_{\pm},\gamma;z] $ and $ U[\alpha_{\pm},\gamma;z] $, 
so the general solution can be written as 
$ w_{\pm} (z) = A_1 F[\alpha_{\pm},\gamma;z] + A_2  U[\alpha_{\pm},\gamma;z] $.
Here, $ F $ and $ U $ are the first and second kind of confluent heypergeometric
functions, respectively. 
However, $ U $ is not regular at the origin and has to be discarded. 
The requirement of normalizability implies that the solution is acceptable 
if $\alpha_{\pm} $ is negative integer $ \alpha_{\pm} = -\nu, \nu = 0, 1, 2,...$. 
This constraint produces the quantization of the energy: 
\begin{equation}
E_{\pm} = \frac{\hbar v_F \lambda}{2} \sqrt{(2\xi_0)^2 - 
(2\xi_0 - (2 \nu  + 1 \pm 1))^2}. 
\end{equation}
The energy eigenvalues are then conveniently written as
\be \label{energy}
E_n = \hbar v_F \lambda \sqrt{(\xi_0)^2 - 
(\xi_0 - n)^2}.
\ee
For $ E_+ $, $ \nu = n - 1 $ and for $E_- $, $ \nu = n$. 
For $ n = 0 $ and $ E_+$, $ \nu = -1 $, but $ \nu $ can not be negative.
For $ n = 0 $ and $ E_-$, $ \nu = 0 $. Therefore, the $ n = 0 $ state is
not degenerate.
For $ n = 0 $, the solution $ w_+ $ does not exist. We will consistently 
incorporate this fact by defining $ w_{n-1} = 0 $.  
The corresponding wave function is given by
\begin{equation}
w_{\pm}(2 \xi) = {}_1F_1[-(n - \frac{1}{2} \mp \frac{1}{2}), 2\beta + 1, 2 \xi],
\end{equation}
where $  {}_1F_1[-n,\alpha, x] $ is the confluent hypergeometric function.

Equation (\ref{energy}) can be re-written as 
\begin{equation}
E_n^2 = \hbar^2 \omega_c^2(x_0) 2n (1 - \frac{n }{2\xi_0}),
\end{equation}
where the local cyclotron frequency is $ \omega_c(x_0) = (v_F)/l(x_0) $.
Note that when $ \bar{\lambda} \rightarrow 0 $ ($\xi_0 \rightarrow \infty $),
the energy eigenvalues reduce to the well known relativistic Landau level 
structure for uniform magnetic field: $ E_n = (\hbar v_F/l(x_0)) \sqrt{2n} $.

When $ \bar{\lambda} $ is very large, the effect of inhomogeneous magnetic 
field vanishes and the Dirac electron feels no effective potential. It 
behaves like a free particle. In this limit, the energy spectrum becomes
$ E_n \propto  n $ which agrees with the known result
for the free Dirac electron \cite{free}.

The complete normalized wavefunctions can be written as
\begin{equation} \label{wf1}
\Psi_{+}^{(n)} = \frac{e^{ik_y y}}{\sqrt{2 L_y l(x_0)}} (2\xi_0)^{\beta} 
e^{-\bar{\lambda} \beta X} e^{-\xi_0 e^{- \bar{\lambda} X}} w_{n-1}
\end{equation}
and 
\begin{equation} \label{wf2}
\Psi_{-}^{(n)} = \frac{ i c_n e^{ik_y y}}{\sqrt{2 L_y l(x_0)}} (2\xi_0)^{\beta} 
e^{-\bar{\lambda} \beta X} e^{-\xi_0 e^{-\bar{\lambda} X}} w_{n},
\end{equation}
where $ L_y $ is the length of the system along the $y$ axis, 
$ c_0 = \sqrt{2} $ and $ c_{n>0} = 1 $. Also, 
$ X = (x - x_0)/l(x_0), \bar{\lambda} = \lambda l(x_0) $, and 
$ w_n $ is given by
\be
w_n  =  \sqrt{\frac{\Gamma[2 \beta + n + 1] \bar{\lambda}}
{\Gamma[2 \beta  + 1] \Gamma[2 \beta ] \Gamma[n + 1]}}
{}_1F_1[-n, 2\beta + 1, 2 \xi_0 e^{-\bar{\lambda} X}].
\ee
We have also checked that the wavefunctions (\ref{wf1}) and (\ref{wf2}) 
reduce to that of the Dirac Landau level for constant magnetic field 
({\it i.e.} a Hermite polynomial multiplied with a Gaussian factor) when 
$ \bar{\lambda} \rightarrow 0 $. The phase factor $ i $ in the lower 
component wavefunction (\ref{wf2}) is obtained from equation (\ref{dir2}), 
which is crucial for calculating the probability current density.

\begin{figure}[ht]
\includegraphics[width=8.0cm]{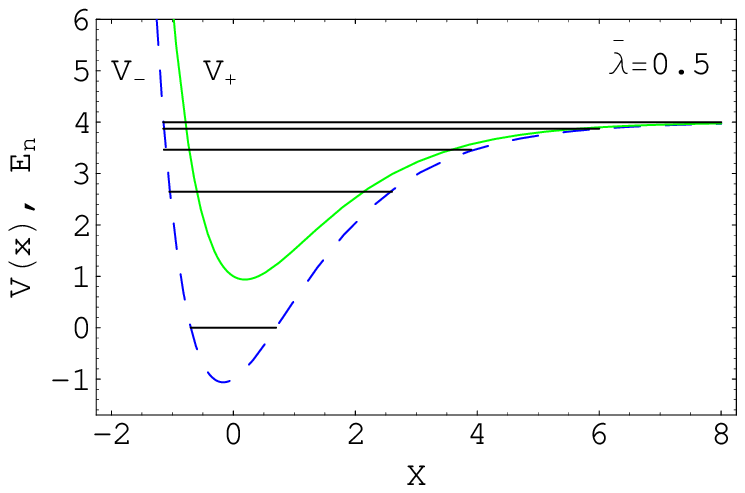}
\includegraphics[width=8.0cm]{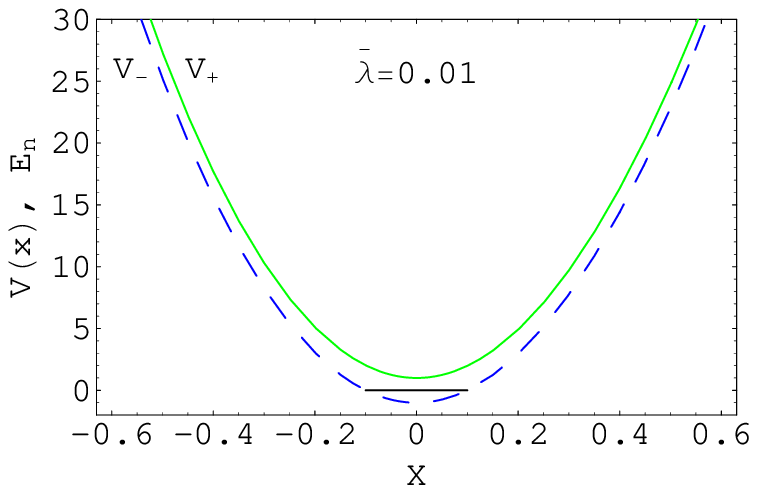}
\caption{(Color online) Plots of the effective potential
$V_{\pm}(x) $ (in units of ($\hbar \omega_c)^2 $) vs $ X $ together with the
energy eigenvalues $ E_n $ (in units of $ \hbar \omega_c $) for
$\bar{\lambda} = 0.5 $ and $\bar{\lambda} = 0.01 $ (only zero-energy level is shown in
this case).}
\end{figure}

The effective potential (\ref{potential}) can be re-written as
\be
V_{\pm} = (\hbar \omega_c(x_0))^2 [\pm e^{-\bar{\lambda} X} +
\xi_0(1-e^{-\bar{\lambda} X})^2].
\ee
For weak-inhomogeneity ($\bar{\lambda} << 1$), the effective potentials
are almost symmetric around $ X=0 $ point.
For strong inhomogeneity ($\bar{\lambda} >> 0$), there is a strong 
asymmetry of the effective potentials.
The effective potentials for negative $k_y$ does not have any global 
minimum. Therefore, the bound state does not exist for negative $ k_y $. 
The effective potentials together 
with the energy eigenvalues for various values of $ \bar{\lambda} $ are shown 
in figure 1. Both the effective potentials get saturated to 
$ (\hbar \omega_c(x_0) )^2 \xi_0 $ at large $X$. The zero-energy state 
($n=0$) is always lies inside the potential $ V_- $ but outside the potential 
$ V_+ $. However, all other discrete energy levels ($ n>0 $) are lying inside 
both the potentials $ V_{\pm} $, which is expected from the solution of the 
Dirac-Weyl equation. The number of energy levels decreases as we increases 
$ \bar{\lambda} $. For example, there are five energy levels including zero-energy 
state when $\bar{\lambda} = 0.5$. On the other hand, number of energy levels 
is quite large when $ \bar{\lambda} = 0.01$. There are a finite number of 
discrete energy 
levels for a given asymmetric parameter $ \bar{\lambda} $. 
The total number of discrete energy levels including zero-energy level 
can be calculated easily from the condition that $ E_n^2 \leq V_{\pm}(X 
\rightarrow \infty ) $ and 
it is given by $ N = Int[\xi_0] + 1 $. Here, $ Int[b] $ means the integer
that is just smaller than $b$.
In figure 1, only the zero-energy level is shown for the 
$ \bar{\lambda} = 0.01 $ case.

The probability density distribution of Dirac electrons in
a given level $ n $ is $ \rho_n (x) = \Psi^{(n)\dag}(x) \Psi^{(n)}(x) $.
The probability density distributions $ \rho_n $ for different values of
$ n $ and $ \bar{\lambda} $ are shown in figure 2.
\begin{figure}[ht]
\includegraphics[width=8.0cm]{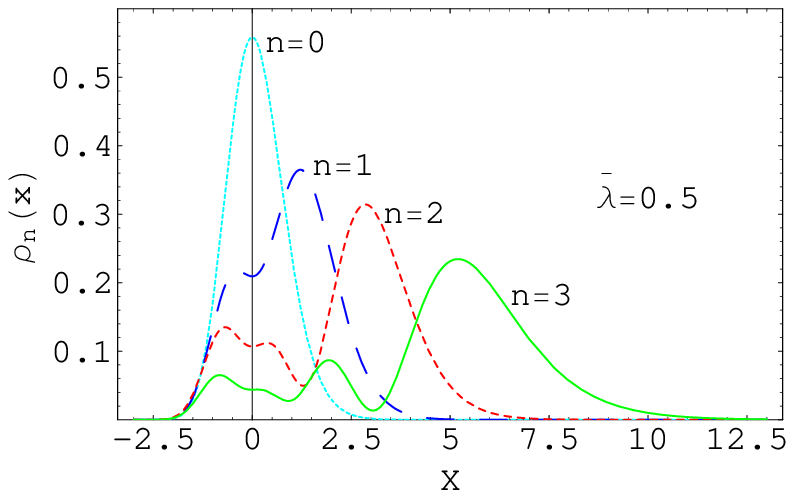}
\includegraphics[width=8.0cm]{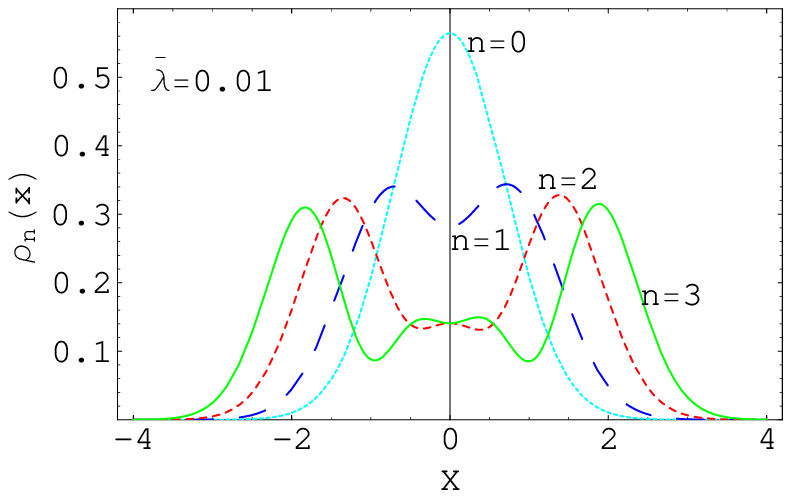}
\caption{(Color online) Plots of the probability density distribution $ \rho_n(x) $
(in units of $ 1/(L_yl(x_0)) $
vs $ X $ for various values of $ n $ with $ \bar{\lambda} = 0.5 $ and $ \bar{\lambda} 
= 0.01 $.}
\end{figure}
The velocity operator that follows from the Heisenberg equation is given 
by $ {\bf v} = v_F {\bf \sigma} $.
The probability current distribution is 
$ J_n(x) = e v_F \Psi^{(n) \dag}(x) \sigma_y \Psi^{(n)}(x)
= -i e v_F (\Psi_{+}^{(n)*}(x) \Psi_{-}^{(n)}(x) - \Psi_{-}^{(n)*}(x) 
\Psi_{+}^{(n)}(x))$. 
The probability current distributions for various values of $ n $ and 
$ \bar{\lambda} $ are  shown in figure 3.
The zero-energy state does not carry any
current, irrespective of nature of the magnetic field.
When $ \bar{\lambda} \rightarrow 0 $, $ \rho_n $ and $ J_n $ are symmetric
around the point $ X=0 $. When $ \bar{\lambda} >> 0 $, $ \rho_n $ and $ J_n $ are
strongly asymmetry as is expected from the effective potentials.

\begin{figure}[ht]
\includegraphics[width=8.0cm]{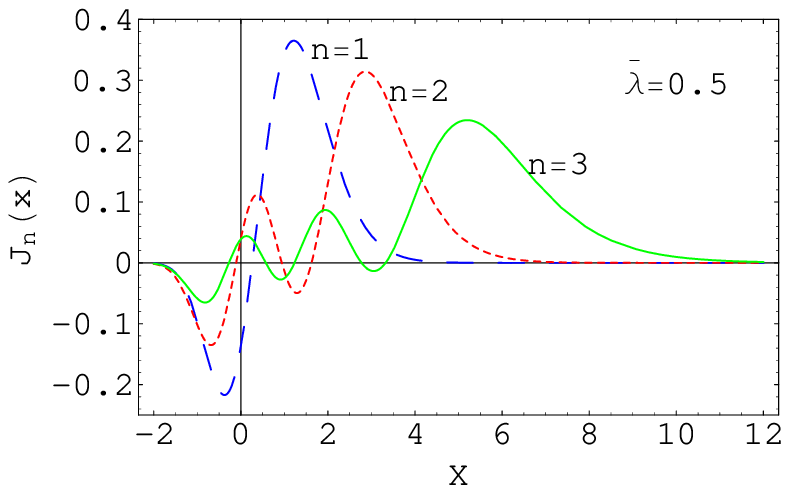}
\includegraphics[width=8.0cm]{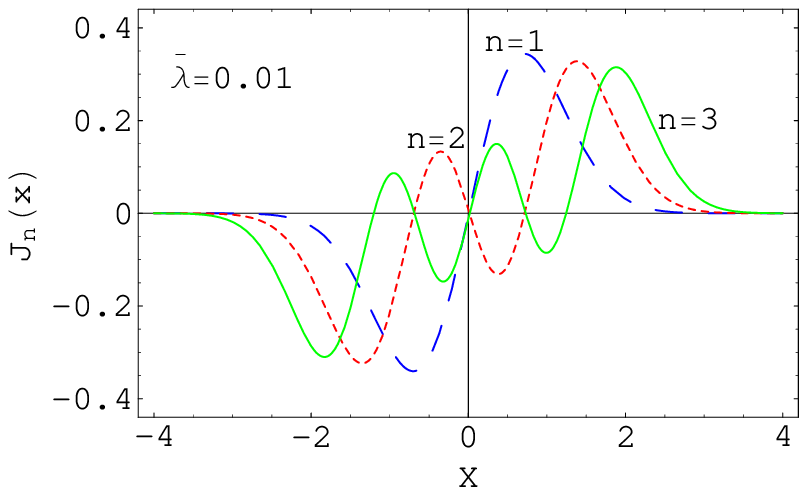}
\caption{(Color online) Plots of the probability current density 
$J_n(x) $ (in units of $ ev_F/(L_y l(x_0) $) vs $ X $ for various values of $n $ 
with fixed $\bar{\lambda} = 0.5 $ and $\bar{\lambda} = 0.01 $.}
\end{figure}

In summary, we have obtained exact energy eigenvalues and the 
corresponding eigenfunctions 
for a Dirac electron in the presence of an exponentially decaying magnetic 
field.
We have also provided the probability density and current distributions for
each band. 

I would like to thank K. Bhattacharya for a discussion.
This work was supported by a research grant (Grant No.: IITK/PHY/20080036) 
of DORD, IIT-Kanpur, India.

\end{document}